# Crime and punishment in scientific research

**Mathieu Bouville**
**<mathieu.bouville@gmail.com>**

Typical arguments against scientific misconduct generally fail to support current policies on research fraud: they may not prove wrong what is usually considered research misconduct and they tend to make wrong things that are not normally seen as scientific fraud, in particular honest errors. I also point out that sanctions are not consistent with the reasons why scientific fraud is supposed to be wrong either. Moreover honestly seeking truth should not be contrived as a moral rule — it is instead a necessary condition for work to qualify as scientific.

I. INTRODUCTION: FRAUD, FROM SCIENCE TO BUREAUCRACY

Both philosophers (Hofmann, 2007; Kaposy, 2008; Schmaus, 1983, 1984) and sociologists (Merton, 1942; Wunderlich, 1974; Zuckerman, 1977, 1984) have asked, in their different ways, what is acceptable scientific behavior. Punishment is not their main concern (even though it may follow from other considerations). On the other hand, scientists, politicians, and lawyers have focused on sanctions for misconduct.[1] There is little interaction between the two: David Guston (1999) found that the changing public policies of the past few decades corresponded to equally changing underlying conceptions of scientific norms. But policy makers did not always notice that a different policy likely means a different construal of scientific fraud; in fact certain policies may not correspond to any coherent theory at all. Policies are essentially *ad hoc*: inclusion or exclusion of certain deeds as scientific misconduct does not come from a careful and consistent argumentation.

To James Dooley and Helen Kerch (2000), "scientific misconduct includes fabrication, falsification, and plagiarism (FFP) of concepts, data or ideas; some institutions in the United States have *expanded* this concept to include 'other serious deviations (OSD) from accepted research practice' " (emphasis added). 'Misconduct' simply means bad behavior, which is quite general. It thus seems that scientific misconduct should be anything that is wrong, not just fabrication, falsification, and plagiarism. In other words, it is the limitation of misconduct to three kinds of acts that requires justification.[2] But whether to include "other serious deviations" is not seen as making a decision regarding what is wrong but rather as a matter of what would make for a more enforceable rule. Dooley and Kerch indeed note that "CRI [the U.S. Commission on Research Integrity] endorsed what it claimed to be a more legally enforceable definition of scientific misconduct" and "more in line with the nation's legal system than with the practice of science." Jesse Goldner (1998) notes that "the general trend over the past decade has been away from a primarily scientist-run process toward one in which lawyers have played an increasingly crucial role." In fact, most of the literature is dedicated to the history of the bureaucracy and of the judicial processes related to research misconduct (e.g. Dooley and Kerch, 2000; Goldner, 1998; LaFollette, 1994; Mishkin, 1999; Pascal, 1999; Price, 1994) rather than to questions of what is wrong and why — the focus is on procedures (due process, whistle-blower protection) and sanctions. (These are not intrinsically legal issues, they have been made so in the U.S. in part because about anything there ends up decided by a tribunal.) There has been little effort to justify prosecuting certain deeds rather than others (many articles offer no justification at all). However, one cannot choose a practical definition of fraud (i.e. a list of banned practices) independently of why fraud is supposed to be wrong. The underlying problem is that when one takes something to be obviously wrong one tends not to scrutinize one's arguments too closely (Bouville, 2008a). The cartoon recently published by *Nature* (vol. 453, pp. 980–981) seemed targeted towards elementary school children rather than scholars: falsification is not nice and you must tell a grown-up if you see someone do it. The present article, on the other hand, aims at a grown readership.

---

1 This trend was triggered by a number of scandals at the end of the last century, which were supposedly indicative of a failure of the self-regulation of science. (Of course, this brand new problem was old news sixty years ago: "violations of professional ethics on the part of scientists are frequent and familiar to all scientists" (Pigman and Carmichael, 1950).) But saying that such scandals show a failure of researchers to fight fraud is like saying that the police seizing a ton of cocaine is proof of its failure to fight drug trafficking. The fact that fraud is found out is not proof that science is not (or no longer) self-regulating: fraud being found out *is* self-regulation.

Research fraud is often presented as a waste of public money, and "surely the public has a right to presume that its tax money is being spent wisely" (Redman and Caplan, 2005). (There is a more general trend making funding central, e.g. to retention and promotion — rather than funding being a means to do science, it seems that science is turning into a means to secure funding.) One should note that a government cannot fire researchers (other than those it directly employs), it cannot prevent anyone from writing in journals or from attending conferences; the only thing it can do is cut funds. The reason why fraud is punished by funding debarment is not that those who sinned by misuse of funds shall perish by funding deprivation, but simply because this is about the only way a government can concretely punish researchers — short of making research misconduct a crime, which only a few (e.g. Redman and Caplan, 2005; Sovacool, 2005) would endorse. Waste of public money is not a serious argument against scientific fraud (as I shall show in greater detail in the next section) but rather a rationalization, a way of justifying the power of investigation of funding agencies. The idea that research misconduct shall be an administrative and legalistic issue precedes the question of why it is wrong.

---

2 The phrases 'research fraud', 'research misconduct', 'scientific fraud' and 'scientific misconduct' tend to be used rather interchangeably, even though 'misconduct' is *a priori* broader than 'fraud.' 'Misconduct' nonetheless seems favored —a search for 'scientific misconduct' turned ten times as many papers as 'scientific fraud'—, possibly because it is vaguer and thus looks more innocuous (concealing the irritating fact that 'scientific fraud' is not an oxymoron). But whether to distinguish between the two or not cannot be a purely semantic issue: it involves a choice as to what should be prosecuted and what should not, i.e. I would get ahead of myself if I made such a decision now. For this reason I will use these four phrases interchangeably. I will also treat 'researcher' and 'scientist' as synonyms.

## II. WHICH PRACTICES ARE WRONG AND WHY?

### A. Why is scientific fraud wrong?

The simplest argument against fraud is that it can "directly and negatively impact the integrity of the research record" ([Dooley and Kerch, 2000](#)), i.e. "of what is presented as scientific facts" ([Forsman, 1999](#)). Research is supposed to create new knowledge, so what is published must be true: scientific journals are not supposed to publish fiction. Another issue is that scientific fraud is incompatible with the necessary "moral integrity of scientists" ([Merton, 1942](#)), and more specifically with "the institutional goal of extending certified knowledge" ([Zuckerman, 1984](#)). Scientists are supposed to honestly seek truth and to aim at creating new knowledge. Claude [Bernard (1865)](#) even talks of "the cult owed to the truth" ("le culte dû à la vérité") and Charles [Babbage (1830)](#) writes: "I feel that I shall deserve the thanks of all who really value truth, by stating some of the methods of deceiving practised by unworthy claimants for its honours." It is often mentioned that scientific misconduct can harm the population; for instance, a fraudulent report that a given medicine has no side effect may expose those taking it to unexpected harm. A last point is that fraud allows some to receive more than they deserve (e.g. promotion or awards) by increasing the number or the impact of their publications. I will scrutinize these five arguments of scientific record, honesty, waste of funds, harm to the population, and undeserved rewards to determine exactly what kinds of deeds they can show to be wrong and to find out whether they have undesirable consequences (e.g. making too many things wrong).

### B. Incriminated practices

The three big players of research fraud are fabrication, falsification and plagiarism. The first two are similar in that they correspond to the publication of results that intentionally misrepresent reality. About all criteria for the wrongness of research misconduct conclude that fabrication and falsification are wrong. However, it is not obvious that they waste funding: someone who fails to obtain interesting results and falsifies data to boost their value does not waste any more funds than someone who fails to obtain interesting results and decides not to publish anything. It is not fabrication and falsification that waste money, but rather the unsuccessful research that led to them. If they waste money, it can only be indirectly: others will unsuccessfully try to build upon fake results, thus wasting funds. But this applies mostly to the higher-profile cases: few papers attract enough attention to have such negative consequences. Whether fabrication and falsification waste money can only be decided case by case.[3]

Plagiarism means passing off someone else's work as one's own. Plagiarism affects our knowledge of who did what rather than our knowledge of the world (the subject matter of science), making it less incompatible with scientific standards than the publication of inaccurate results: "the introduction of fraudulent evidence is more seriously at odds with the institutional goal of extending certified knowledge than is the publication of plagiarized but reliable evidence" (Zuckerman, 1984). In fact plagiarism can even help to spread valid information by multiplying sources. Plagiarism does not use any money: plagiarism is the absence of work, it takes no time and no resource (it is precisely this absence of work that makes it wrong). So *a fortiori* plagiarism does not waste any public funds. Plagiarism does not harm the population at large since the results are correct. The main impact of plagiarism is that plagiarists will be overestimated because their publication list is unduly impressive, which skews promotion and award procedures.[4] (One should nonetheless note that fraud can bias promotion and awards only if it is massive: one is unlikely to receive an award for one fraudulent paper in a low-profile journal.)

---

[3] Among the prosecuted cases of fabrication or falsification, those that wasted money are overrepresented because it is generally this unfruitful attempt at repeating the experiments that created suspicion. In other words, fabricated and falsified results that waste money are prosecuted more often than others because they are more visible, not because they are more wrong. Even though this correlation between prosecution and waste of money is accidental, it is not problematic if waste of funding is the reason why fabrication and falsification are wrong. Otherwise one should prosecute all fraud (even if it leads to no wasteful work by others) — current practices are at odds with arguments other than that of waste of funds. Another situation where the probability of prosecution may not be related to wrongness is popular work — a report for *Science* magazine "urges us to give special attention to a relatively small number of papers that are likely to be especially visible or influential" ([Kennedy, 2006](#)). One would be more likely to be prosecuted if one's work is more visible even though visibility plainly has nothing to do with wrongness (and, unlike in the previous case, this would be intentional).

Other practices that are often frowned upon are redundant publications, 'self-plagiarism,' and 'salami slicing' (i.e. publishing many papers where one would have been enough), as well as unwarranted ('honorary') authorship. Another problem relative to publications is when reviewers stall the review of an article (by not replying to the editors in a timely manner or by asking for extra work that will take time and delay publication) so that their own paper on the subject will come out first. The main issue is, as with plagiarism, that one may receive undeserved rewards.

Certain results may be obtained from human subjects who had not consented to being part of an experiment or from animals that were mistreated in the process. One should note that mistreating subjects does not affect the truthfulness of the research and of publications, neither does it harm the population at large or waste money. As Chris Kaposy (2008) notes: "Of course, accurate objective scientific results can be derived from experiments that pay no heed to patient safety or other ethical issues. It is not a necessary truth that objective knowledge must be produced ethically." Mistreatments, mishandling of hazardous materials and similar issues are simply a separate question and one must argue against it using specific arguments.

C. Side effects

After looking at whether arguments against scientific misconduct justify banning certain practices, I now turn to the question of whether they make collateral damages, e.g. whether they make wrong things that are generally not so considered. Table I provides a summary.

An important issue is whether false results coming from an intent to deceive are to be treated differently from those that are due to involuntary errors, carelessness or chance. Honest errors do not mean that the scientist was not honestly seeking truth, but they do adversely affect scientific records, waste funds, and they may harm the population or get someone undeserved rewards. In other words, most arguments should hold honest errors to be as wrong as fabrication or falsification. Honest errors may in fact be more problematic, simply because they are far more common than voluntary deception (Bouville, 2008b) — science is hurt more by "sloppy or careless research practices and apathy than by incidents of research fraud or other serious scientific misconduct" ([Institute of Medicine](#), [1989](#), p. 21). One should note that *any* error can have negative consequences: not only the intentional, not even only the reckless or the negligent: even an error made in spite of extraordinary caution can damage the scientific record, waste money, hurt the population, and garner undeserved rewards. Prosecuting the unlucky is then as justified as prosecuting the dishonest — surprisingly, this straightforward and important conclusion is seldom reached. It is commonly claimed that cheaters cannot be trusted, and "the institution of science involves an implicit social contract between scientists that each can depend on the trustworthiness of the rest" ([Zuckerman](#), [1977](#), p. 113). But, one could just as much say that someone who is incompetent or careless cannot be trusted because future results may be tainted by these flaws. Both technical incompetence and dishonesty can break trust. Moreover, loss of trust is to some extent self-fulfilling (Bouville, 2008b): one loses trust because one believes that fraud can lead to loss of trust.

---

[4] *Nature* (2006) laments that some countries "offer scientists cash prizes for publications in top-level international journals" because "a researcher measuring science in terms of dollars might be more tempted to plagiarize or fabricate data." (One may notice that the same is true of someone measuring science in terms of publication in *Nature*.) Since the main effect of plagiarism is undeserved rewards —fame, money—, one may wonder why plagiarism is wrong if seeking fame or money is wrong, i.e. if the victims are deprived of something that they should not want anyway. In fact greed and vanity are taken into account to manage researchers through promotion and awards: someone absolutely free from such vices would be absolutely unmanageable. Russian mathematician Grigori Perelman declined a Fields medal (the 'Nobel prize for math') in 2006, saying that it "was completely irrelevant for [him]. Everybody understood that if the proof is correct then no other recognition is needed." Strangely, he was not praised for his utter absence of vanity and his unalloyed search for knowledge.

The harm to the population argument applies essentially to medical research (fake cosmology data will not harm anyone, for instance), so it cannot be used against research misconduct in general — it would mean that some fields would be necessarily devoid of misconduct. One should also note that one study should not be able to cause much harm, since something that may cause much harm should receive much attention (verification, repetition), so that a work that can single-handedly cause much harm can do so only through the actions or inaction of others (i.e. this cannot be single-handed).

In some cases the wrongness of fraud does not justify trying to reduce it. This is obvious in the case of the wasting of funding: investigation and prosecution of fraud can be more costly than fraud itself. These trade-offs will be more fully addressed in the next section.

A last point is that if something is wrong because it has certain negative consequences then anything with the same consequences (within or without science) is wrong. A government cutting health care budgets does more harm to public health than all scientific fraud combined. Likewise, scientific misconduct is a tiny source of loss of public funding: if efficient use of public funds is really the issue then one should look at billions rather than at spare change. The war on Iraq has cost the U.S. trillions of dollars ([Stiglitz and Bilmes](#), [2008](#)) and not even the most faithful believer can deny that there have been many mistakes, costing billions of dollars and many human lives. If one really means to save public funds then what really wastes them should be the priority, not scientific misconduct. One should also bear in mind that research is intrinsically a very wasteful and inefficient activity: success is always preceded by many shots in the dark and many failures — misconduct is just a drop in this ocean.

D. The ground of the arguments

The five arguments of scientific record, honesty, waste of funds, harm to the population and undeserved rewards mostly fail to support current policies on research fraud: they may not prove wrong what is typically considered research misconduct and they tend to make wrong things that are not usually seen as scientific fraud, such as honest errors. But one must also ask whether we should give any weight to these arguments at all.

While it is plain that science is a social activity, arguments such as waste of money and undeserved rewards go one step further by construing science as just an expense or a source of social recognition. An obvious problem is that this makes science an occupation similar to any other, without much special to it: science is just a job. (One may remark that the concept of disinterestedness introduced by Robert [Merton](#) ([1942](#)) was meant for researchers *qua* scientists, whereas nowadays they are rather treated *qua* employees: 'publish or perish' is a matter of management, not of science.) Furthermore, this would make scientific fraud not a matter of science but rather a financial or managerial issue — but then why should scientists care about it at all?

The question of honesty is not self-evident either. First, how does one know that fraud actually springs from dishonesty (Bouville, 2007c)? Friedman (1992) pointed out the lack of "study that gets inside the mind of the perpetrators to discover what, if anything, they were thinking when they committed fraud" (the only such work is that of Davis *et al.*, 2007) and Forsman (1999) argued that "in protecting research integrity, we have to face the real human psychology." Another point is that if one holds that honesty is necessary to reach the truth then it is just a means (i.e. honesty is necessary inasmuch as one cannot generate knowledge without it), but this is just another version of the scientific record argument rather than a separate viewpoint. One may alternatively say that honesty is required for its own sake, but this requires justification.

### E. Examples

After looking at which acts are wrong according to various arguments against research misconduct, I will now consider a number of scientists and compare them based on the same arguments. Table II gives a summary.

Author A copied an article *verbatim* and published it in a marginal journal, where nobody read it. Author B is both honest and careful but nevertheless made an error and thus published results that mistakenly hold a certain drug (now used by millions) to be harmless; this work made B famous. Unlike A, B published results that are false, wasted money, harmed people and garnered undeserved rewards. A is guilty of scientific misconduct.

Author C did not feel like doing actual work and simply made stuff up, the ensuing article had no impact. The research of D failed to generate any result, so D did not publish anything. Neither C nor D harmed anyone or received unmerited distinctions; D wasted money while C did not. C is guilty of scientific misconduct.

One can also compare researchers B and D. Neither was dishonest but both wasted funds, and scientist B published invalid results, harmed the population, and received undeserved rewards. Neither researcher would be considered guilty of scientific misconduct.

E is the serious researcher next door: E's articles are honest, essentially error-free and of moderate interest. F copied a fairly recent paper that went unnoticed in spite of its great merits; these merits were finally recognized (and attributed to F) and led to very promising new developments. Neither E nor F published false data, only F's article was of great benefit to science and to the population at large (and at absolutely no cost). Arguably, F should be rewarded for furthering the goals of science. F is guilty of scientific misconduct.

Author G (who is not a native speaker of English) oftentimes copies paragraphs from the background and method sections of other people's articles in order to ensure clarity of these descriptions (the methods themselves are well-established); G's results are genuine and they have had very beneficial outcomes. G is not trying to steal from others but simply to improve the clarity of his articles (see Bouville, 2008b). The few paragraphs G copied cost no money, garnered no unwarranted reward, and did not affect the scientific record. G's research had positive consequences for the population. G is guilty of scientific misconduct.

One can also compare public relation (PR) plagiarist F to non-native speaker G. Neither scientist wasted funds, published invalid results, or harmed the population. F was dishonest but not G. Both researchers would be considered guilty of scientific misconduct.

H gave an untested drug to patients without their consent; it turns out that this medicine did not harm them and is even the only known cure to a certain disease. The only difference between E and H is that H's research was greatly beneficial to the population, unlike that of E. H's behavior would be considered wrong (but H may not be prosecuted).

There is a dramatic difference between current policies and the outcomes of the usual arguments against fraud, as can be seen in Table II — no single argument against research misconduct seems to capture the many aspects (positive and negative) of the behaviors of these scientists and of their consequences (and neither do policies): they may discriminate between researchers that exhibit no major difference, and they may even punish the one who benefited science or the whole of society.

III. SANCTIONS

### A. The purpose of sanctions (if any)

As Philip Ball (2008) points out, "austere calls for penalizing scientific misconduct rarely indicate what such penalties are meant to achieve", because "discussions of scientific misconduct seem all too often to stop at the primitive notion that it is a bad thing." For instance, "the South Korean national committee on bioethics rejected an application by Hwang Woo-suk to resume research on stem cells. Why? Because 'he engaged in unethical and wrongful acts in the past', according to one source." Ball wonders: "Does the committee fear that Hwang would do it again, despite the intense scrutiny that would be given to his every move? Do they think he hasn't been sufficiently punished yet? Or perhaps that approval would have raised doubts about the rigour of the country's bioethics procedures?"

It is not clear that sanctions have a goal and that those deciding them would be able to justify punishment beyond saying that the scientist did something wrong. Bernard Williams (1985, p. 177) noted that "blame is the characteristic reaction of the morality system" and Friedrich Nietzsche (1888) that "wherever responsibilities are sought, it is usually the instinct of wanting to judge and punish which is at work." The First World Conference on Research Integrity held in September 2007 in Portugal used the picture of a hand-cuffed Erlenmeyer flask as a logo of sorts (see http://tinyurl.com/2odvre). Barbara Redman and Arthur Caplan (2005) entitled their article "Off with their heads!". Birgitta Forsman (1999), who believes that "the most important thing is not to punish scientists who have done something wrong," must feel lonely. Punishing fraudsters seems so natural to so many that no one seems to notice that punishment always needs justification — "The moral problem that the having of a legal institution of punishment presents can be stated in one sentence: It involves the deliberate and intentional infliction of suffering" (Burgh, 1982).

B. Consistency and relevance of sanctions

Since sanctions are the main concern, discussions revolve around what is prosecutable rather than around what is wrong. It may be impossible to prove intent but possible to show that someone's actions were at least reckless. Punishing recklessness may then be used as a tactic to punish those who intended to deceive but could not be proven so — "because proving intent is very difficult, the addition of 'committed…in reckless disregard of accepted practices' was identified as an improvement in the policy" (Bird and Dustira, 2000). But the fact that intent is hard to prove does not entail that one should include reckless errors along with intentional ones. One should prosecute people because they did something wrong, not because it is easy to prosecute them. Recklessness should be treated as wrong only if it is wrong (and if it is believed to be wrong, one must provide arguments in this sense). What kind of justice does one get when one prosecutes based on convenience rather than based on wrongness?

Just like a given argument against scientific misconduct can make certain behaviors wrong but not others, it can justify certain kinds of sanctions but not others. One may wonder why funding agencies should debar a plagiarist from funding, since plagiarism does not waste their money, it does not affect the creation of new knowledge, and it has no adverse effect on the population at large — put colloquially: it's none of their business. If a plagiarist received an undeserved promotion or if an award was obtained for fabricated data, these may be rescinded. But if the plagiarist would have been promoted anyway (e.g. his other work would justify promotion) then, in order to cancel the promotion —let alone fire the researcher or rescind his PhD ([Nature](), [2004]())—, one must claim that plagiarism is intrinsically wrong and that plagiarists should be sanctioned in any case, i.e. one must invoke an argument other than that of undeserved promotion. If the argument is 'without the fraudulent articles they would not have been promoted' then the promotion can be canceled (based on this argument) only if it is true that they would not have been promoted in the absence of the fraudulent papers.

### C. Consequences and fairness of sanctions

What is the concrete effect of a one-year debarment from funding? None for someone about to take a sabbatical leave, who has retired, or a then-graduate student who no longer does research — the impact of the sanction will depend greatly on the specific situation of the researcher, i.e. on things utterly irrelevant to what the scientist did wrong. Also, consequences are not proportional to the length of the debarment: someone without funding for five years must find a new occupation; in such a case the end of the sanction is purely theoretical. One, three, five years cannot be treated just as numbers: they have particular effects on scientists, and it is these consequences of the sanctions rather than the length of the debarment that should be proportional to the wrongness of what the researcher did. However, the only available study (Redman and Merz, 2008) does not have this level of detail and the U.S. Office of Research Integrity [ORI] does not know the concrete consequences of fund debarment of various lengths (ORI, private communication). In other words, no one knows whether the harshness of the actual punishment depends on the wrongness of the deed.

Certain consequences of a condemnation can randomly increase the sanction. As [Babbage]() ([1830]()) notes, "that part of the scientific world whose opinion is of most weight, is generally so unreasonable, as to neglect altogether the observations of those in whom they have, on any occasion, discovered traces of the artist" — if guilt is proven for one article it is assumed for all others (also see [Nature](), [2004]()). Moreover, each time someone writes on the subject of scientific fraud, decade-old deeds are once more associated with the name of a researcher so that "the potential negative impact on reputation [is] perpetuated (and hence punishment extended)" even though one could think that "someone who has made a mistake should be allowed the opportunity to rehabilitate his or her reputation" ([Bird](), [2004]()). (Not to mention the largely publicized 'Baltimore case,' thus named because David Baltimore was not the one accused.)

### D. On sanctioning negligence

Warren [Schmaus]() ([1983]()) takes negligence to be wrong because it springs from a failure to follow proper research procedures, in particular it is a lack of scrutiny of one's own work. The negligent researcher is not aware of the potential error (but should have been), whereas the reckless is aware of it and fails to act upon this knowledge. A practical problem is that identification of negligence will be influenced by hindsight bias. Indeed, when an error is uncovered, everyone will claim *in retrospect* that obviously the scientist should have spotted it — similar to sport fans who always know after the match what tactic to use. There are hundreds of things which may go wrong (and which the researcher is expected to prevent) but when it is known exactly what went wrong, this one thing tends to leap to the top of the list of what should have been tested. Negligence is thus likely to be overdiagnosed. And, contrary to the difficulty to show an intention to deceive, this artifact is naturally against the defendant, i.e. against presumption of innocence.

On the bright side, negligent errors are easily avoided: all one needs to do is check for possible artifacts and contaminations, repeat the experiments many times, etc. — all this takes is time and money, which are both plentiful. Naturally if one spends more time and money checking results, one spends less generating new findings. Whether having fewer, more reliable results is better than more numerous but less reliable results is an open question. The ideal situation is probably enough verification to weed out the most egregious errors but not so much as to do nothing but check results. In other words, it is good that some results are invalid. (Thermodynamics tells us that a state devoid of defect is seldom the most favorable: entropy should be reduced only inasmuch as this decreases free energy.) Science would not fare better if all negligent work were sanctioned, since it would force researchers to dedicate far more resources to checking results. If any error that I could have detected can lead to loss of funding, infamy, and so on, I will indeed be more careful. Actually, I will be careful to the point of meaninglessness. (And I could be accused of wasting money on unnecessary tests.) Creativity, uncertainty, and even errors are part of scientific research and any attempt to get rid of every error would obviously put an end to any scientific endeavor.

In fact, we may not need the same level of accuracy for all results. If I publish data that turn out to be of no interest to anyone then whether they are valid or not is rather unimportant — an error would affect a volume of the scientific record that no one ever browses anyway. If, on the other hand, my results attract lots of interest, many people will try to reproduce them, so one will soon find out about possible errors. The most scrupulous verifications should be saved for results that deserve them. This would be a more efficient use of time and money than painstakingly checking all results. A process that would attempt to remove every error at every stage would be highly inefficient.

### E. When fighting fraud is worse than fraud

Fighting fraud may have negative effects similar to those of fraud itself. If fraud is wrong because it wastes money then fighting it makes sense only if this saves money. If hiring new specialized employees instead of new postdocs and spending less time on research and more on paperwork costs more than fraud then fighting fraud is a waste of public money and should be fought. If fraud were actually a financial issue, then one would make sure that fighting fraud be not more costly than fraud itself. Firms have quality controls because very low quality costs them money and is bad publicity but, since overquality could cost more money than low quality, a reasonable trade-off is sought.

Similarly, the obsession with fraud may hinder science even more than fraud itself. Kenneth Pimple (1999) does not "fear that ORI (as it is currently constituted) will *intentionally* hamper science, but [he] know[s] that bureaucracies always cause trouble, even when they do not mean to." Dooley and Kerch (2000) note that "most researchers in the physical sciences are likely subject to a policy on scientific misconduct but are unaware of what would actually happen if they were accused of misconduct." Physicists should feel concerned only if they are likely to exclaim 'Someone must have been telling lies about me, I have done nothing wrong but, one morning, I was arrested.' If this is unlikely to occur then awareness is unneeded. And if this is a likely event where you live, learning about potential consequences of accusations of misconduct is not the answer: exile is. Sovacool (2005) wants to "penalize those who know of misconduct within their institution, harshening criminal statutes could motivate more colleagues to report violations" and notes that "criminalizing misconduct could also motivate scientists to be more careful in their research." In fact it would most likely "transform science so that … well, so that it isn't fun any more" (Pimple, 1999). It would not motivate scientists but rather disgust them and incite them to find a different occupation. Orwellian science is an oxymoron, and should remain so.

One should remark that if certain talented researchers are banished from science then science may suffer a great loss (also see Ball, 2008). If Einstein had been found guilty of some misdemeanor in his youth, would it have been a good idea to prevent him from doing science any longer? It seems that the harm to science would have been incommensurably greater than would have been the lack of a punishment. A scientist who would publish great work alternating with fraud would still contribute far more to science than someone publishing honest but infinitely uninteresting work.

IV. CONCLUSION: FRAUD, FROM MORALITY TO SCIENCE

Zuckerman (1984) notes that negligent researchers are treated "with derision and contempt" whereas "scientists respond to [fraud] with all the sting of moral indignation, denouncing it as a crime and labeling perpetrators as charlatans and scoundrels." In other words: in science, morality is far more important than science itself. A corollary is that for something to be important it has to be contrived as a moral requirement. One can thus find a purely technical recommendation such as "Act with skill and care in all scientific work. Maintain up to date skills and assist their development in others." as part of a proposed "universal ethical code for scientists" (Council for Science and Technology, 2005). Kaposy (2008) notes that "According to these codes of ethics, it is an ethical duty to be skillful and rigorous in pursuit of the truth — it is an ethical duty to do the science well." Hofmann (2007) claims that "moral norms are not only needed to regulate science, they define it", and Schmaus (1983) takes moral concerns to have a special power when he focuses on "a professional function upon which society places a high moral value", and since knowledge is not a "high moral value" (as it is not a moral value at all) it is dismissed as unimportant. Seeking the truth is presented as a matter of morality —as honesty— instead of being related to knowledge and understanding, i.e. the core purposes of science. It is somehow taken for granted that if seeking the truth were just a scientific matter it would be negligible, but as an ethical issue it becomes all important (also see Bouville, 2007b). An obvious consequence is that the more one looks at science and knowledge from a moral viewpoint, the more one belittles them by denying them any intrinsic value.

In fact, the difference between fabricating results and obtaining false data by error is not that the former is immoral. Rather the difference is that fabricating data is not science whereas involuntary errors are part of science (perhaps not the most commendable part, but science nonetheless). There is no need to import external criteria: science comes with its own. What is scientific is the most natural and best criterion. This has some notable consequences. (i) If it is established that some work is not scientific then withdrawing the article is quite natural. However, this concerns only the work not its author: there is a leap between saying that the work is not scientific and saying anything about the author, e.g. that he is not a real scientist. While not being scientific seems the most natural way of rejecting certain works, it is not the criterion that most straightforwardly leads to punishment (which may explain why it is seldom mentioned). (ii) Let us consider the case of a researcher who faked results in one paper but who otherwise published genuine scientific work. Most people would consider that this researcher should be deprived of funding and may even be dismissed. Now let us consider a scientist who also writes on subjects such as parapsychology or astrology. Again part of his publications would not be scientific. Would one think of depriving him from funding or even of his job? (iii) What if a physicist who has been caught fabricating data claims that it is not fraud but really an attempt to deconstruct the status quo of the bourgeois patriarchy? Naturally, other physicists would reject this as non-sense. Broader institutions, on the other hand, may be in trouble: can universities employ (and public money fund) people who aim at doing just this while fighting fraud? Also, honestly seeking knowledge is not universally seen as a goal; some research may be censored because some do not want to know the answer to certain questions (e.g. Ceci *et al.*, 1985) — convenient beliefs may be preferred to inconvenient knowledge. Naturally, all this is problematic for institutions, not for scientists. Scientists need only realize that there is no reason to allow the issue of fraud to be treated as a bureaucratic or a moral matter. Zuckerman (1984) noted that "scientists respond to [fraud] with all the sting of moral indignation" — in fact this indignation should be of a different kind, a scientific one.

TABLE I Which acts are wrong according to current policies and to various arguments against research misconduct. In bold are cases conflicting with current policies. *: these are held to be wrong but may not be prosecuted as 'scientific misconduct' (rules may vary by country, etc.). †: there is a correlation between prosecution and waste of funds (see footnote 3).

|  | current policies | scientific record | honesty | waste of funds | harm to population | undeserved rewards |
|---|---|---|---|---|---|---|
| construes science as | — | knowledge | moral | financial | serviceable | managed |
| fabrication | yes† | yes | yes | some† | some | some |
| falsification | yes† | yes | yes | some† | some | some |
| plagiarism | yes | **no** | mildly | **no** | **no** | some |
| redundant publication | yes* | **no** | **no** | no | no | some |
| 'honorary' authorship | yes* | **no** | **no** | no | no | some |
| stalling reviews | yes* | **no** | mildly | no | no | some |
| mistreatments | yes* | **no** | **no** | **no** | **no** | **no** |
| honest errors | no | **yes** | no | **yes** | some | some |
| all fields | yes | yes | yes | yes | **no** | yes |
| trade-off | — | **yes** | no | **yes** | **yes** | no |
| outside science | — | no | no | **yes** | **yes** | no |

TABLE II Which researcher did the *worse* thing according to current policies and to various arguments against research misconduct. In bold are cases conflicting with current policies. 'Neither' and 'both' respectively mean that neither researcher and both researchers did something wrong. *: mistreatment of human subjects is held to be wrong but may not be prosecuted as 'scientific misconduct' (rules may vary by country, etc.). E†: it is not that E did something bad, rather others did something good, making E the worse of the two.

|  | current policies | scientific record | honesty | waste of funds | harm to population | undeserved rewards |
|---|---|---|---|---|---|---|
| A (plagiarist) vs. B (unlucky hecatomb) | A | **B** | A | **B** | **B** | **B** |
| C (lazy fabricator) vs. D (unsuccessful) | C | C | C | **D** | **neither** | **neither** |
| B (unlucky hecatomb) vs. D (unsuccessful) | neither | **B** | neither | **both** | **B** | **B** |
| E (average) vs. F (PR plagiarist) | F | **E†** | F | **E†** | **E†** | arguable |
| E (average) vs. G (non-native speaker) | G | **neither** | **neither** | **neither** | **E†** | **neither** |
| F (PR plagiarist) vs. G (non-native speaker) | both | **neither** | F | **neither** | **neither** | arguable |
| E (average) vs. H (no consent) | H* | **neither** | **neither** | **neither** | **E†** | **neither** |

13